\documentclass[11pt, letterpaper]{article}

\usepackage[T1]{fontenc}
\usepackage[utf8]{inputenc}
\usepackage{mathpazo}          
\usepackage[scaled=0.88]{helvet} 
\usepackage{microtype}

\usepackage[
  top=1in, bottom=1in, 
  left=1in, right=1in,
  headheight=14pt
]{geometry}
\usepackage{setspace}
\usepackage{amsmath, amsfonts, amssymb}
\usepackage{graphicx}
\usepackage{natbib}
\usepackage{url}
\usepackage{hyperref}
\usepackage{xcolor}
\usepackage{float}
\usepackage{wrapfig}
\usepackage{tikz}
\usetikzlibrary{positioning, arrows.meta, calc, backgrounds, decorations.pathreplacing}
\usepackage{fancyhdr}
\usepackage{titlesec}
\usepackage{enumitem}
\usepackage{etoolbox}

\definecolor{accent}{HTML}{1E3A5F}
\definecolor{rulecolor}{HTML}{C8CDD3}
\definecolor{bodytext}{HTML}{23272F}
\definecolor{captiontext}{HTML}{4B5563}
\definecolor{linkblue}{HTML}{1D4ED8}

\definecolor{fg}{HTML}{1a1a1a}
\definecolor{mg}{HTML}{555555}
\definecolor{lg}{HTML}{999999}
\definecolor{vlg}{HTML}{cccccc}
\definecolor{ambcol}{HTML}{3B82F6}
\definecolor{ambfill}{HTML}{EFF6FF}
\definecolor{ambline}{HTML}{BFDBFE}
\definecolor{percol}{HTML}{0D9488}
\definecolor{outcol}{HTML}{DC2626}
\definecolor{outfill}{HTML}{FEF2F2}
\definecolor{outbord}{HTML}{FECACA}

\color{bodytext}

\hypersetup{
    colorlinks=true,
    linkcolor=linkblue,
    citecolor=accent,
    urlcolor=linkblue
}

\titleformat{\section}
  {\large\bfseries\sffamily\color{accent}}
  {\thesection}{0.7em}{}
\titleformat{\subsection}
  {\normalsize\bfseries\sffamily\color{accent}}
  {\thesubsection}{0.6em}{}
\titlespacing*{\section}{0pt}{1.8em}{0.6em}
\titlespacing*{\subsection}{0pt}{1.4em}{0.4em}

\renewcommand{\headrulewidth}{0.4pt}
\renewcommand{\headrule}{\hbox to\headwidth{\color{rulecolor}\leaders\hrule height \headrulewidth\hfill}}
\fancypagestyle{plain}{%
  \fancyhf{}
  \fancyfoot[C]{\small\sffamily\color{captiontext}\thepage}
  \renewcommand{\headrulewidth}{0pt}
}

\usepackage[font={small,sf}, labelfont={bf,color=accent}, textfont={color=captiontext}]{caption}

\makeatletter
\renewcommand{\maketitle}{%
  \thispagestyle{plain}
  \begin{center}
    {\color{rulecolor}\rule{\textwidth}{0.5pt}}\\[1.2em]
    {\LARGE\bfseries\sffamily\color{accent} The Epidemiology of Artificial Intelligence\par}
    \vspace{0.9em}
    {\color{rulecolor}\rule{\textwidth}{0.5pt}}\\[1.4em]
    {\normalsize
      Harsh Parikh\textsuperscript{1}\,,\;
      Tyler McCormick\textsuperscript{2}\,,\;
      Emily Johnson\textsuperscript{3}\,,\;\\
      Leo Hickey\textsuperscript{4}\,,\;
      Megan Ranney\textsuperscript{1}\,,\;
      Bhramar Mukherjee\textsuperscript{1}\par}
    \vspace{0.5em}
    {\small\itshape\color{captiontext}
      \textsuperscript{1}Yale University \quad 
      \textsuperscript{2}University of Washington \quad
      \textsuperscript{3}University of Southern Denmark \quad
      \textsuperscript{4}Vassar College \quad
      \par}
    \vspace{1.2em}
  \end{center}
}
\makeatother

\renewenvironment{abstract}{%
  \begin{center}
  \begin{minipage}{0.92\textwidth}
  \vspace{0.5em}
  {\color{rulecolor}\hrule height 0.3pt}
  \vspace{0.8em}
  {\small\sffamily\bfseries\color{accent} Abstract}\\[0.4em]
  \small\setstretch{1.35}%
}{%
  \vspace{0.8em}
  {\color{rulecolor}\hrule height 0.3pt}
  \end{minipage}
  \end{center}
  \vspace{1em}
}

\begin{document}

\maketitle

\begin{abstract}
Artificial intelligence (AI) systems increasingly shape how people access health information, make medical decisions, and receive care---yet epidemiology lacks frameworks for measuring AI exposure or studying its health effects at the population level. Here we argue that AI now functions as a determinant of health and propose a conceptual framework, borrowed from environmental epidemiology, for studying it. We distinguish \emph{ambient AI exposure}---algorithmic curation and AI-mediated institutional decisions that affect populations regardless of individual choice---from \emph{personal AI exposure}---direct, volitional use of AI tools. We characterize AI's possible causal roles in epidemiological models, show that existing experimental approaches are inadequate for capturing chronic, population-level effects, and illustrate these ideas with nationally representative US survey data. We discuss implications for study design, health equity, and AI governance.
\end{abstract}


Artificial intelligence has moved from research laboratories into the infrastructure of daily life. ChatGPT reached 100 million users within two months of its launch \citep{hu2023chatgpt}; by 2025, one in six American adults uses AI chatbots for health information at least monthly \citep{kff2024aiuse}, two-thirds of US teenagers have used them \citep{pew2025teens}, and two-thirds of US physicians report using AI in clinical practice \citep{ama2024physician}. In early 2026, OpenAI launched ChatGPT Health, integrating users' medical records and wearable data for personalized guidance \citep{openaihealth2026}, and other major platforms followed with similar health-focused agents \citep{googlehealth2026}. AI is no longer a specialist tool---it is embedded in the everyday infrastructure through which health decisions are made and care is delivered. This is not entirely new: algorithmic prediction models have become standard in clinical risk stratification, insurance underwriting, and resource allocation \citep{obermeyer2019bias}. But generative AI---systems that produce novel content and interact with users in real time---has increased the scale, visibility, and directness of AI's contact with populations.

This embedding generates benefits and harms simultaneously. AI helps patients translate medical terminology and prepare informed questions \citep{pew2024aihealth}; AI-powered chatbots extend mental health support to underserved populations \citep{fda2025dhac,merrill2025chatbots}; AI-based triage improves emergency department prioritization \citep{taylor2025aied,hinson2025triage}. Conversely, companion chatbots foster emotional dependency in adolescents \citep{garcia2024characterai,namvarpour2025teen}, and students who outsource cognitive tasks to AI show reduced neural connectivity \citep{kosmyna2025brainonllm}. As AI grows more adept at mimicking humans, the picture complicates further: AI-generated health misinformation is indistinguishable from human-written content \citep{spitale2023ai,zhou2023synthetic}, and AI-produced arguments shift health attitudes as effectively as human persuasion \citep{bai2025artificial}.

Both point to the same conclusion: AI shapes \emph{what} health decisions people make, \emph{how} they make them, and---through clinical tools and algorithmic triage \citep{taylor2025aied,dacosta2025aitriage}---\emph{what care they receive}. Yet epidemiology lacks a framework for conceptualizing, measuring, or studying these effects. We have established protocols for air pollution, nutrition, tobacco, and social media---but not for AI.

In this Perspective, we argue that \emph{artificial intelligence is a determinant of health that demands systematic, population-level study.} We show why existing experimental approaches are inadequate, then propose a framework---borrowed from environmental epidemiology---that distinguishes ambient from personal AI exposure, describes AI's causal roles, and outlines the dimensions along which exposure must be characterized. We illustrate these ideas with nationally representative US survey data \citep{amerispeak2025ai}. Our aim is to propose a research agenda, not to settle one.

\section{Micro-experiments miss the population perspective}

A growing literature treats AI as an exposure, but within narrow experimental paradigms. In one line of work, AI-generated text serves as a persuasive intervention: AI-written policy arguments shift attitudes as effectively as human ones \citep{bai2025artificial}; large language models produce misinformation indistinguishable from human text \citep{spitale2023ai}; AI-generated COVID-19 misinformation uses distinct linguistic strategies that evade detection \citep{zhou2023synthetic}. In another, interaction with AI is the exposure: co-writing with a biased model shifts authors' own attitudes \citep{jakesch2023co}; biased AI search suggestions move voting preferences \citep{epstein2024search}; repeated ChatGPT reliance reduces neural connectivity \citep{kosmyna2025brainonllm}. In clinical contexts, algorithms trained on historical cost data systematically underestimate illness severity in Black patients \citep{obermeyer2019bias}, and large language models exhibit covert dialect-based biases in triage recommendations \citep{hofmann2024covert}.

These studies share a limitation: they define AI exposure as a low-dimensional treatment, delivered in the context of a specific task, with outcomes measured over short horizons. This is akin to studying diet by measuring a single meal in a laboratory---informative about mechanisms but unable to capture the health consequences of sustained dietary patterns. A further complication is that the evidence base lags the technology: the most rigorous studies were conducted on earlier-generation models whose capabilities have since been surpassed, meaning the exposure landscape shifts even as we characterize its effects. Key public health questions remain unanswered: What are the effects of \emph{chronic} AI use on depression and cognition? How does sustained reliance on AI for health advice change care-seeking behaviour? Do shifts in platform governance alter mental health outcomes at a population scale?

Answering these questions demands a population-level epidemiological paradigm: defining AI exposure as a multidimensional, time-varying construct; measuring it in cohorts; linking it to longitudinal health outcomes; and deploying study designs---prospective cohorts, target trial emulation, quasi-experiments exploiting policy and version variation---suited to this complexity \citep{hernan2020causal,robins2000msm}. The framework we propose supports that transition.

\section{A framework for AI exposure and health}

\subsection{AI as an algorithmic determinant of health}

We adopt a pragmatic definition: AI refers to \emph{technology that enables computers to simulate human learning, comprehension, problem-solving, decision-making, and creativity} \citep{ibmai2024}---encompassing both predictive systems that risk-stratify or allocate resources, and generative agents that produce novel content and interact with users in natural language. The former often operates as ambient exposure (shaping institutional decisions); the latter typically constitutes personal exposure (individuals engaging directly). Both share an inherent opacity that makes it difficult for users and policymakers to assess when to trust their outputs.

Public health has long recognized that health is shaped by biological, environmental, behavioural, and social determinants \citep{lalonde1974new,marmot2005social}. Recent work extends this to ``digital determinants of health'' (DDoH): van Kessel et al.\ identified 127 health determinants that emerged or changed during the digital transformation \citep{vanKessel2025ddoh}; the WHO European Region catalogued digital factors shaping health outcomes across member states \citep{who2024digitalfactors}; and Hatef et al.\ proposed a framework integrating DDoH into health equity research, emphasizing digital access, literacy, and infrastructure as upstream influences \citep{hatef2025framework}.

AI is not merely another digital technology to catalogue alongside broadband access and electronic health records. It is \emph{causally active} in a way that motivates specific epidemiological treatment. What sets AI apart is the conjunction of four properties not found together in other exposures. First, AI \emph{generates novel content}: unlike broadband or EHRs, which transmit existing information, AI creates text, images, and recommendations---simultaneously the medium and the message. Second, AI \emph{adapts to the individual in real time}, creating feedback loops between exposure and behaviour that static technologies cannot. Third, AI \emph{operates across multiple determinant categories simultaneously}---reshaping information environments, modifying health behaviours, and reconfiguring social relationships. Fourth, the exposure is \emph{non-stationary and non-transparent}: model updates alter the ``agent'' between observations, and the design of chatbots to mimic humans makes it difficult for individuals to recognize when they are being changed by AI.

We describe this cross-cutting, adaptive, generative, and non-stationary exposure as an \textbf{algorithmic determinant of health}---not to claim it constitutes a clean fifth category in a formal taxonomy, but to signal that existing frameworks, including the DDoH literature, do not yet provide the causal and measurement tools needed to study it. The van Kessel and Hatef frameworks \citep{vanKessel2025ddoh,hatef2025framework} catalogue \emph{what} digital factors affect health; we are concerned with \emph{how to study their effects}---what the exposure is, how to measure it, and what causal roles it can play.

\subsection{Ambient and personal AI exposure}

To make AI exposure tractable for epidemiological analysis, we borrow a distinction from environmental epidemiology: \emph{ambient} versus \emph{personal} exposure \citep{bell2004pollution}.

In air pollution research, ambient exposure is the concentration of pollutants in the shared environment---shaped by industrial emissions, traffic, geography, and regulation. Personal exposure is an individual's actual intake, depending on behaviour, location, and physiology. The two are related but not identical: people in the same city experience different doses. Crucially, the framework is \emph{policy-aware}: different levers operate at different levels---individual, family, community, state, national, and global. AI creates an analogous structure, with one important difference: unlike air pollution, AI can both help and harm human health.

\begin{wrapfigure}{r}{0.52\textwidth}
\centering
\vspace{-0.5em}
\includegraphics[width=0.50\textwidth]{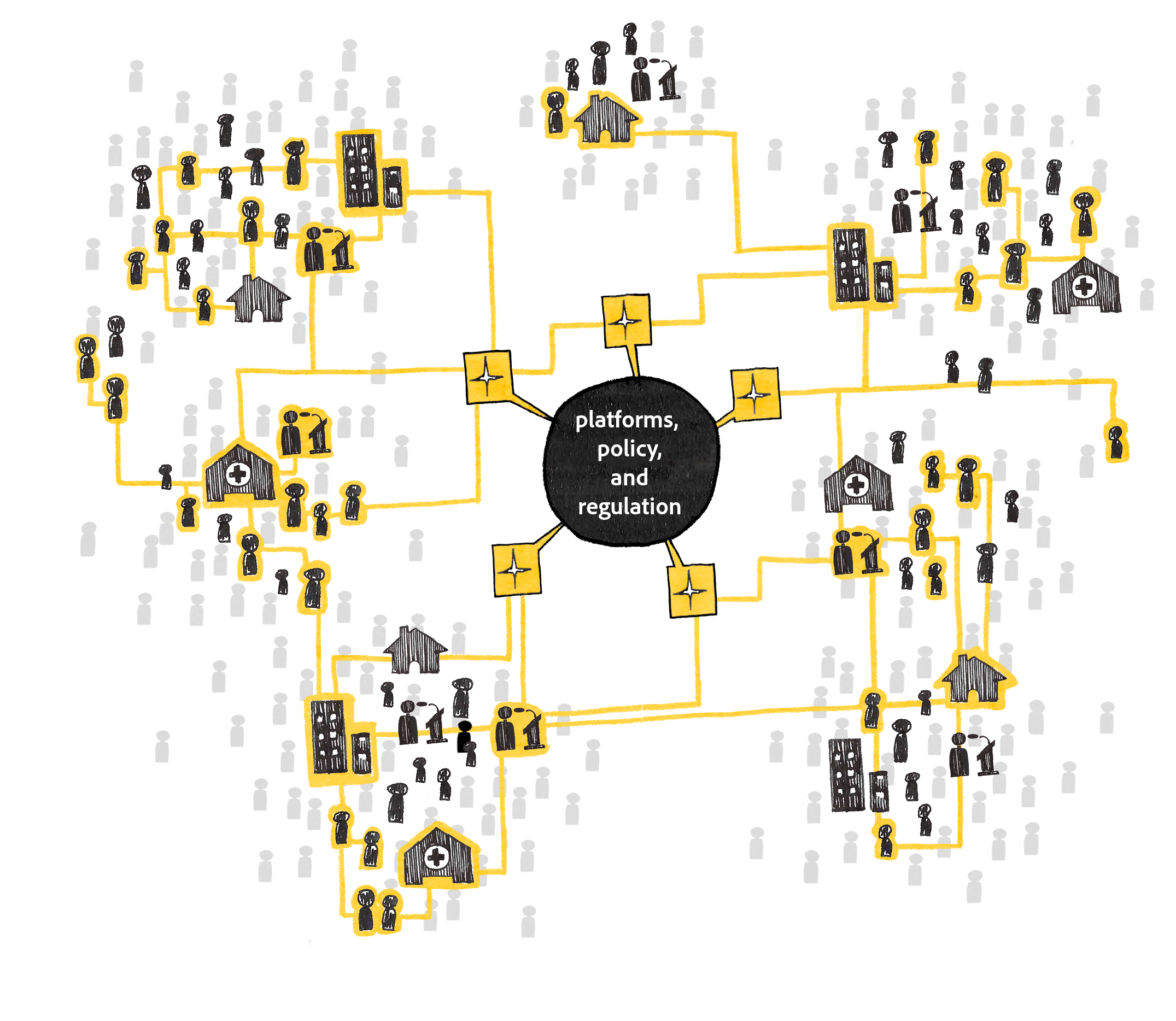}
\caption{AI exposure as a societal phenomenon. Individuals inhabit an environment where institutions (buildings) and AI systems (gold squares) mediate interactions, governed by a central regulatory layer. Individuals with gold silhouettes directly and volitionally use AI---a form of \emph{personal AI exposure}. Individuals with black silhouettes do not use AI directly but are nevertheless exposed through their social networks or the institutions they interact with---a form of \emph{ambient AI exposure} (gold pathways). Grey silhouettes represent individuals not exposed to AI---neither directly nor ambiently. ``Not using AI'' $\neq$ ``not exposed to AI.''}
\label{fig:illustration_ambient}
\vspace{-0.5em}
\end{wrapfigure}

\paragraph*{Ambient AI exposure} refers to the shared algorithmic layer of the information and institutional environment. Algorithms curate news feeds, rank search results, triage emergency department patients, score credit applications, and moderate online content---affecting populations \emph{regardless of whether individuals choose to use AI directly}. A person who has never used ChatGPT may still have their insurance claim adjudicated by an AI, their social media feed shaped by recommendation algorithms, or their emergency wait time set by algorithmic triage \citep{obermeyer2019bias,taylor2025aied}. Beyond clinical settings, the World Bank uses AI to estimate welfare and poverty in data-scarce settings, with these estimates feeding geographic targeting and programmatic decisions \citep{worldbank2022poverty,worldbank2023social}; WHO and the Global Fund use AI models to quantify local malaria incidence and populations at risk \citep{who2022metadata,globalfund2025results}.

Ambient AI exposure is shared, largely invisible, and shaped by institutional decisions rather than individual choice. Unlike traditional environmental exposures tied to geography, this layer is \emph{non-local}: a single model update can alter exposures for millions regardless of their coordinates. AI exposure also propagates through social networks---a challenge familiar from infectious disease and network epidemiology.
\paragraph*{Personal AI exposure} refers to direct, volitional interactions with AI systems: consulting a chatbot for health advice, relying on AI for emotional support, co-writing with a language model, or using a symptom checker. Like personal pollution exposure, it varies by access, preferences, literacy, and purpose. Measuring it will likely require survey data or passively sensed AI use logs.

\begin{wrapfigure}{l}{0.52\textwidth}
\centering
\vspace{-0.5em}
\includegraphics[width=0.50\textwidth]{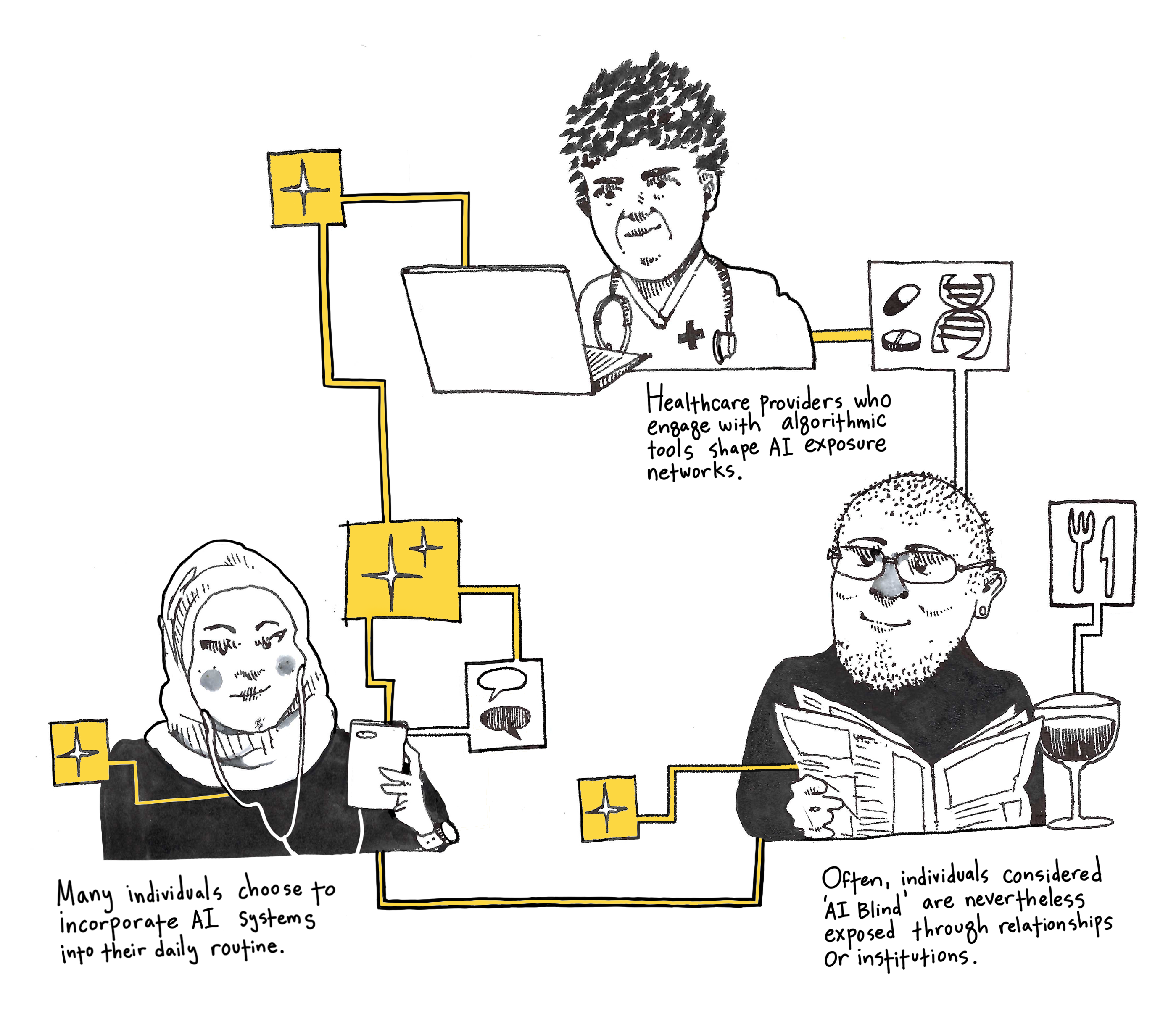}
\caption{Heterogeneity of AI exposure. A healthcare provider (top) uses algorithmic diagnostic tools, shaping the AI exposure of her patient (bottom right)---who reads a newspaper and does not use AI directly but is exposed through the provider's decisions and a family member's AI-mediated care (an ``AI blind'' individual). The caretaker (bottom left) volitionally uses an AI mental health app. These three individuals differ in access, purpose, intensity, and dependency, illustrating why multidimensional measurement is necessary.}
\label{fig:illustration_heterogeneity}
\vspace{-0.5em}
\end{wrapfigure}

This distinction clarifies that ``not using AI'' does not mean ``not exposed to AI.'' It identifies different intervention targets---regulation for ambient exposure, behaviour and education for personal---and maps onto the epidemiological toolkit: ambient exposure supports ecological and quasi-experimental designs; personal exposure supports cohort studies with individual-level measurement. Figures~\ref{fig:illustration_ambient} and~\ref{fig:illustration_heterogeneity} illustrate the framework.

\paragraph*{Scope and limits of the analogy.} We import \emph{methodological} tools from environmental epidemiology---the ambient/personal distinction, dose-response thinking, population surveillance---not to imply that AI is inherently toxic. Unlike air pollution, AI exposure may be beneficial, harmful, or both, depending on context. The dose-response relationship may be non-monotone, with benefits at moderate levels and harms at extremes. But the analogy extends naturally to \emph{dose, chronicity, and susceptibility}: just as pollution epidemiology considers concentration $\times$ duration and vulnerable subpopulations, an epidemiology of AI must reckon with intensity of use, sustained reliance, and populations at heightened risk---adolescents, individuals in mental health crises, patients with low health literacy \citep{pew2025teens,kirk2025parasocial,fang2025longitudinal}.

\subsection{Causal roles of AI in epidemiological models}

Depending on the research question, AI can occupy at least four roles in epidemiological models. As \textbf{exposure}, direct interaction with AI or exposure to AI-generated content affects health: a teenager's daily use of a companion chatbot is a personal exposure whose effects on depression can be studied prospectively \citep{pew2025teens,namvarpour2025teen}; a population's exposure to AI-curated health misinformation is an ambient exposure suited to ecological designs \citep{meyrowitsch2023chatbots}. As \textbf{confounder}, AI algorithms that shape both what content people encounter and downstream outcomes introduce bias: a study of social media use and depression that ignores algorithmic curation suffers from unmeasured AI confounding \citep{epstein2024search}. As \textbf{mediator}, AI transmits upstream determinants: income determines access to premium AI tools, which shapes the quality of health information, which shapes health decisions---placing AI on the causal pathway between socioeconomic status and health \citep{osonuga2025bridging}. As \textbf{effect modifier}, AI alters established relationships: quality AI decision support might buffer the effect of physician inexperience on diagnostic accuracy \citep{taylor2025aied}; AI-generated misinformation might amplify the effect of low health literacy on poor health choices \citep{spitale2023ai,monteith2024ai}. These roles are not mutually exclusive and may shift as systems and behaviours evolve. The mediator and modifier roles, in particular, are hypotheses awaiting direct empirical evidence.

Defining AI’s causal role also requires grappling with violations of the Stable Unit Treatment Value Assumption (SUTVA) that underpin classical epidemiology. AI breaks SUTVA in two ways. First, through \emph{interference}: because algorithms curate shared environments based on collective behaviour, one person’s interaction with an AI system can alter the algorithmic feed of their peers---personal exposure generates ambient exposure for others. Second, through \emph{multiple versions of treatment}: because generative systems adapt to individual users in real time, the ``same’’ nominal exposure---daily use of a specific chatbot---manifests as a uniquely tailored intervention for every person. Future epidemiological models must therefore move beyond assuming independent units, employing network-based causal inference to disentangle the direct, indirect, and total effects of algorithmic deployment.

\paragraph*{A worked example: chronic chatbot use and incident depression.} To illustrate how the framework generates study designs, consider estimating the effect of sustained chatbot use on incident depression among adolescents aged 13--17. \emph{Eligibility}: no prior depression diagnosis. \emph{Exposure}: chatbot use over 12 months, measured via self-report and passively sensed app-usage logs \citep{insel2017digital}. \emph{Outcome}: incident depressive episode (PHQ-A $\geq$ 11 or clinical diagnosis). \emph{Confounders}: baseline mental health, socioeconomic status, social media use, peer relationships. \emph{Causal method}: methods for continuous, time-varying exposure. \emph{Key challenges}: defining ``daily use'' when platforms change; separating chatbot effects from displacement of other activities; selection bias (depression-prone adolescents may seek chatbot support); and non-stationarity (the chatbot itself changes between measurements). The framework generates specific, testable designs.

\subsection{Measuring AI exposure}
 
If AI exposure is to serve as an epidemiological construct, it requires structure. We propose two layers.

At the \textbf{individual level}: (1) \emph{Access}---connectivity, devices, technical capacity \citep{osonuga2025bridging,hatef2025framework}; (2) \emph{Tool portfolio}---which systems, through which channels; (3) \emph{Intensity}---frequency, duration, and trajectory of use; (4) \emph{Purpose}---education, work, health, emotional support, entertainment \citep{pew2025teens,sidoti2025schoolwork}; (5) \emph{Dependency}---degree of reliance and social acceptability in the user's context \citep{namvarpour2025teen,kirk2025parasocial}.

\begin{figure}[!htbp]
\centering
\includegraphics[width=0.88\textwidth]{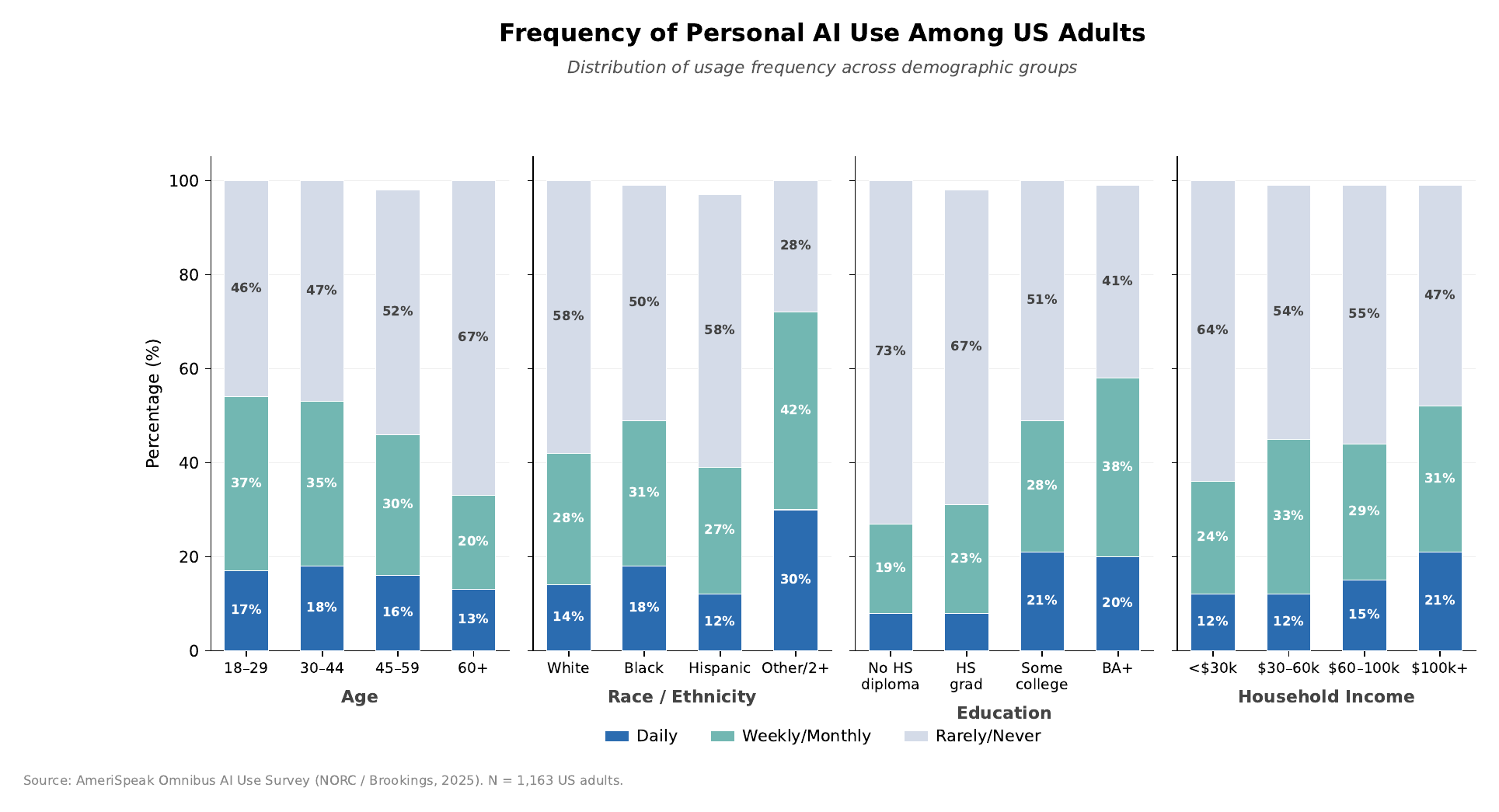}
\caption{Frequency of personal AI use among US adults, illustrating the \emph{intensity} dimension of AI exposure. Adults aged 60+, those with lower educational attainment, and lower-income households are most concentrated in the ``rarely or never'' category. The starkest contrast is at the education threshold: 73\% of those without a high school diploma report rarely or never using AI, compared to 41\% of BA+ holders. Data: AmeriSpeak \citep{amerispeak2025ai}, $N = 1{,}163$.}
\label{fig:frequency}
\end{figure}

\begin{figure}[!htbp]
\centering
\includegraphics[width=0.88\textwidth]{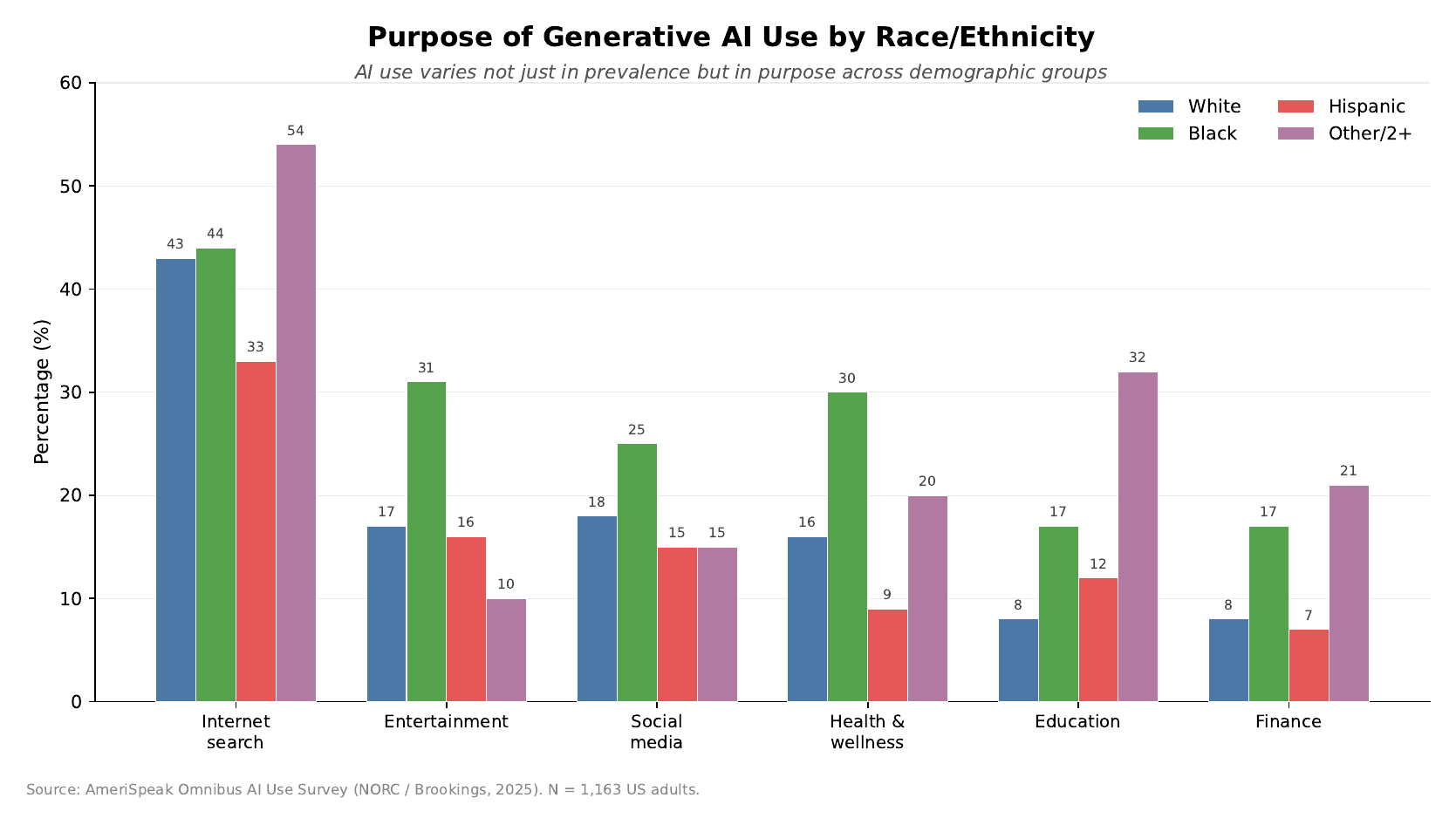}
\caption{Purpose of generative AI use by race/ethnicity, illustrating the \emph{tool portfolio} and \emph{purpose} dimensions. AI use varies not only in prevalence but in function: Black adults report the highest rates of health-related AI use (30\%) and entertainment (31\%); Other/2+ adults lead in internet search (54\%) and education (32\%). These patterns underscore that epidemiological measurement of AI exposure must capture \emph{what} people use AI for, not merely \emph{whether} they use it. Estimates for individual race $\times$ purpose cells are based on small samples and should be treated as suggestive. Data: AmeriSpeak \citep{amerispeak2025ai}, $N = 1{,}163$.}
\label{fig:purpose}
\end{figure}

As illustrated in Figure~\ref{fig:illustration_heterogeneity}, these dimensions generate different epidemiological predictions. The intensity of AI use varies steeply by education: 73\% of adults without a high school diploma rarely or never use AI, compared to 41\% of BA+ holders (Figure~\ref{fig:frequency})---a gradient that parallels the education gradient in health, raising testable hypotheses about AI as a mediator of socioeconomic health disparities. The purpose of use varies by race/ethnicity: Black adults disproportionately use AI for health and wellness, while Other/multiracial adults lead in education and internet search (Figure~\ref{fig:purpose}). If health-related AI use confers benefits or risks, Black adults will experience these effects first---likely reflecting underlying health burden and limited affordability of traditional care rather than simple preference. The same nominal ``AI exposure'' label can conceal fundamentally different experiences differing in frequency, function, and likely health consequences.

At the \textbf{institutional level}: (1) \emph{Regulatory environment}---data-protection laws, AI-specific regulation; (2) \emph{Platform governance}---data retention, safety filters, anonymity guarantees; (3) \emph{Model properties}---training data provenance, bias profiles, update frequency. Two tools identical at the interface may generate very different exposures depending on their training data and regulatory constraints.

A key property of this framework is that AI exposure is \emph{nudgeable}: jointly shaped by individual choices and institutional structures, much as alcohol consumption reflects both preference and taxation. This makes AI a viable target for intervention at both levels---digital literacy on one side, regulation and platform redesign on the other. For epidemiologists, nudgeability implies that policy changes create natural experiments: when a platform modifies its safety filters or a government enacts AI regulation, the resulting exposure shift can be exploited quasi-experimentally to estimate health effects.

\section{Population data reveal a measurement gap}

Most of what we know about AI use comes from platform data---ChatGPT interaction logs, app telemetry, web traffic---that capture only active users, reflect no population denominator, and systematically exclude the unexposed. This is a \emph{selection bias} problem: studying AI's health effects from platform data is like studying smoking's effects using only data from cigarette purchasers.

\begin{figure}[!htbp]
\centering
\includegraphics[width=0.85\textwidth]{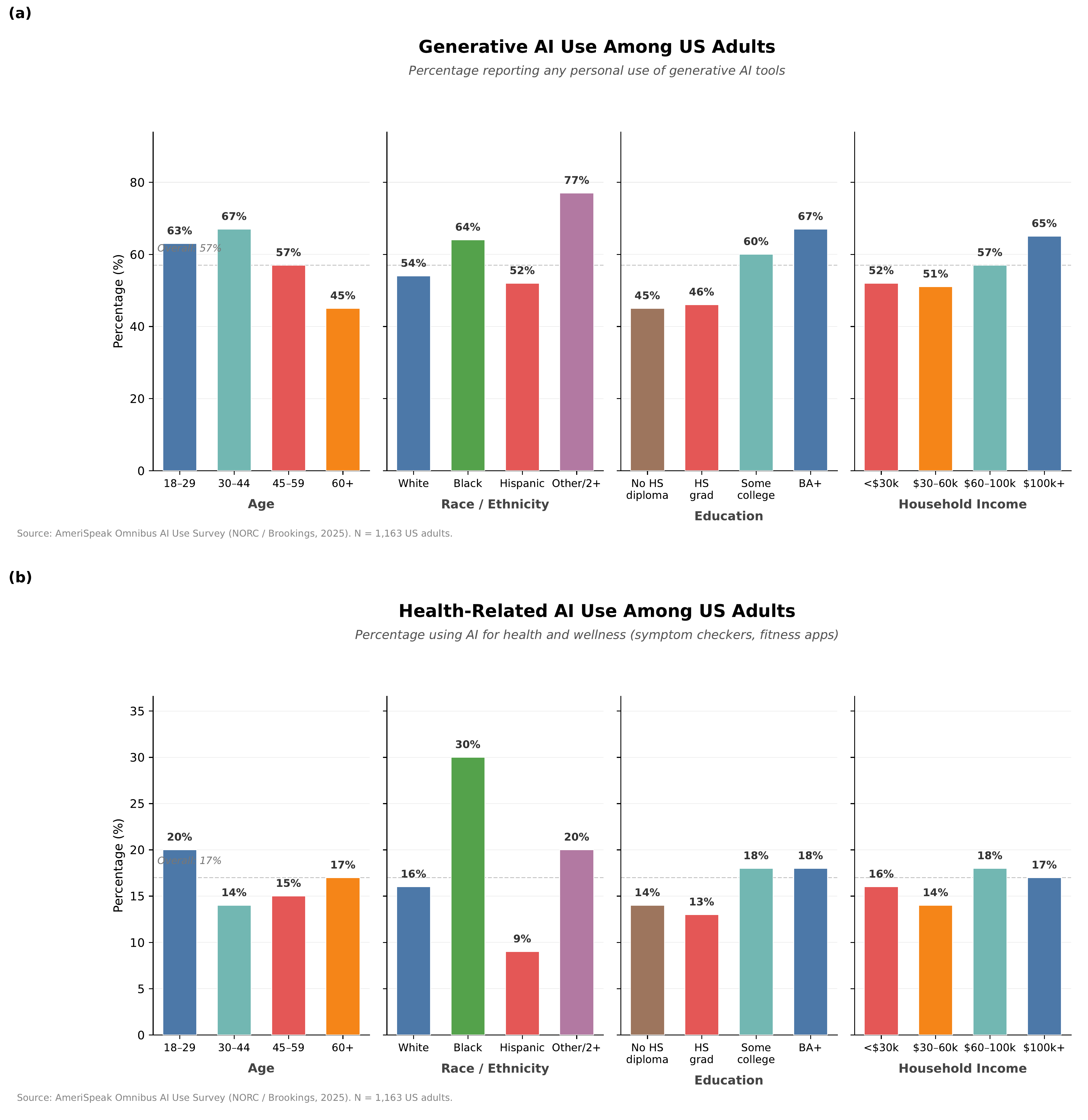}
\caption{AI use among US adults. \textbf{(a)}~Generative AI use by age, race/ethnicity, education, and household income. The dashed line indicates the overall prevalence (57\%). AI adoption varies substantially across demographics: adults aged 30--44, Black adults, those with a bachelor's degree or higher, and those earning \$100k+ report the highest rates. The Other/2+ category ($n = 112$) is heterogeneous, combining Asian, Native American, Pacific Islander, and multiracial respondents; its high point estimate should be interpreted cautiously. \textbf{(b)}~Health-related AI use (symptom checkers, fitness apps). The demographic pattern diverges from overall adoption: Black adults report the highest health-related AI use (30\%), nearly double the overall rate of 17\%, while Hispanic adults report the lowest (9\%). This divergence illustrates that general AI adoption patterns do not predict health-specific exposure---a critical consideration for epidemiological measurement. Subgroup estimates for race/ethnicity are based on small cell sizes and should be interpreted with appropriate caution. Data: AmeriSpeak Omnibus Survey \citep{amerispeak2025ai}, $N = 1{,}163$ US adults.}
\label{fig:anyuse}
\label{fig:healthuse}
\end{figure}

Nationally representative surveys reveal a different picture. The AmeriSpeak Omnibus Survey \citep{amerispeak2025ai} finds that 57\% of US adults use generative AI for at least one personal purpose, but this figure masks substantial variation. Daily use is concentrated among college-educated adults (20--21\%) versus 8\% among those without college education---a step function at the education threshold. Across racial/ethnic groups, daily use ranges from 12\% (Hispanic) to 30\% (Other/multiracial; $n = 112$, interpret cautiously). Health-related AI use shows its own demographic texture: Black adults report the highest rates (30\%) compared to White (16\%) and Hispanic (9\%). Pew surveys reveal similar heterogeneity among adolescents \citep{pew2025teens}. These patterns underscore that AI exposure is not uniform and its health effects are likely heterogeneous (Figure~\ref{fig:anyuse}).

Within our framework, these patterns generate testable predictions. The concentration of health-related AI use among Black adults implies that health effects of AI-mediated information will disproportionately affect this population first. The education gradient implies AI may \emph{mediate} the socioeconomic gradient in health: if frequent use improves health literacy among college-educated adults while remaining inaccessible to others, AI widens existing disparities. Conversely, populations with the highest intensity of use may bear unexpected risks from misinformation exposure or cognitive dependency. These are hypotheses, not foregone conclusions, but they illustrate how the framework converts descriptive patterns into a research programme.

A further challenge is data obsolescence. With each new model release, each new integration into smartphones and wearables, exposure and behaviour patterns shift. Tracking these changes at the population level demands collaboration among AI platforms, users, and survey researchers.

This is a \emph{measurement gap}, distinct from the selection bias above: no existing nationally representative survey links AI use to health outcomes. We can describe who uses AI and how often, but cannot yet study whether that use matters for health at the population level. An epidemiology of AI must attend to both populations: those whose heavy use may generate harm and those whose exclusion from AI constitutes a different kind of deprivation.

\paragraph*{International variation.} Our empirical illustrations draw on US data, but the framework applies globally. The ambient exposure layer differs markedly across regulatory contexts: the EU AI Act mandates transparency for high-risk AI; China requires opt-out mechanisms for recommendation systems; many low- and middle-income countries have no AI-specific regulation. A global epidemiology of AI should exploit this regulatory heterogeneity---cross-national comparisons can help disentangle institutional from individual-level effects. Existing surveillance infrastructure (WHO digital health monitoring, EU digital economy surveys) could be extended to include AI exposure measures at modest incremental cost.

\section{Implications for practice, equity, and governance}

\paragraph*{For epidemiological practice.} AI exposure should be routinely measured in population health studies, as diet, physical activity, and screen time already are. Self-report is a pragmatic starting point but needs calibration: respondents misestimate digital behaviour \citep{ernala2020facebooktime}. Digital phenotyping via smartphone-based passive sensing offers a validation pathway \citep{insel2017digital}. Study designs must accommodate time-varying, adaptive exposures: marginal structural models, g-computation, target trial emulation, and quasi-experimental approaches exploiting policy shifts are well-suited \citep{robins2000msm,hernan2020causal}. Researchers should also account for the ambient component of AI exposure, particularly when comparing outcomes before and after mass AI deployment. In areas where many deaths occur outside the healthcare system, for instance, AI algorithms already impute causes of death, forming the basis for estimates of cause-specific mortality burden \citep{who2022verbal}.

\paragraph*{For health equity.} If AI shapes health, unequal access becomes a health equity concern---but so does \emph{differential harm} \citep{osonuga2025bridging,celi2022sources}. Populations lacking access to quality AI may be disadvantaged; those with the heaviest exposure---adolescents forming parasocial bonds \citep{kirk2025parasocial,fang2025longitudinal}, individuals in crisis using unregulated therapy bots---may bear disproportionate harm. Both deprivation and concentration are equity-relevant.

Algorithms internalize and reproduce the social structures embedded in their training data---what Airoldi terms \emph{machine habitus} \citep{airoldi2021machine}. We see this when clinical algorithms systematically underestimate illness severity in Black patients \citep{obermeyer2019bias}, or when language models exhibit covert biases against non-standard dialects \citep{hofmann2024covert}. The distribution of harms is rarely additive: individuals at the intersection of multiple marginalized identities experience compounded, uniquely structured AI exposures that broad demographic averages conceal \citep{buolamwini2018gender}. Epidemiological measurement must ensure these concentrated harms are not smoothed over in population-level estimates.

The non-locality of ambient AI exposure adds a further dimension. AI models routinely transform fragmented surveillance data into comparable measures of health loss for global priority-setting, and the assumptions they embed shape which populations and risks become visible in policy debates. AI can therefore act as a mechanism for automating inequality \citep{eubanks2018automating,benjamin2019race}---encoding harms that propagate at scale before surveillance can identify them.

\paragraph*{For AI governance.} If AI systems shape population health, the entities operating them bear responsibility for supporting independent research---much as pharmaceutical companies must conduct post-market surveillance. But a disanalogy applies: \textbf{AI companies control the data needed to study their own products.} In environmental and pharmaceutical research, government-mandated reporting gives researchers independent access. For AI, platform telemetry remains proprietary. Closing this gap requires mandated data-sharing, independent auditing, privacy-preserving access protocols, and AI exposure registries analogous to pharmacovigilance networks.

\section{Limitations and open questions}

Several limitations apply. First, the ambient/personal distinction, while analytically useful, may not cleanly partition real-world exposures: a person scrolling algorithmically curated health content is simultaneously subject to ambient curation \emph{and} making a personal choice to engage (Figure~\ref{fig:illustration_heterogeneity}). Second, the causal roles we describe---particularly mediator and effect modifier---are conjectures; direct empirical demonstrations remain scarce. Third, we focus primarily on generative AI, but the ``algorithmic determinant'' claim arguably extends to all algorithmic systems (recommendation engines, automated decision tools, predictive policing), raising questions about scope. Finally, the pollution analogy risks pathologizing AI use. The dose-response relationship is likely non-monotone and context-dependent, and many AI use cases should not be classified as strictly harmful or beneficial. Any regulatory framework derived from this analogy must account for genuine benefits alongside harms.

\section{Conclusion}

AI is already shaping health---through the information people encounter, the advice they follow, and the clinical decisions made on their behalf. We have proposed a framework that distinguishes ambient from personal exposure, identifies causal roles for AI in epidemiological models, and outlines the dimensions along which exposure must be measured. Whether this specific framework proves durable matters less than the underlying claim: AI's health effects deserve the same systematic, population-level scrutiny that epidemiology brings to other environmental and behavioural exposures.

Much remains to be worked out. The right measurement instruments do not yet exist. Longitudinal cohorts linking AI use to health outcomes have not been established. The dose-response relationship is unknown and may not resemble anything in our existing toolkit. The analogy will inevitably break down in places; it is a starting point.

The gap between AI's penetration into health-relevant domains and our ability to study its effects is widening. Closing it will require new data infrastructure, new study designs, and collaboration among public (both AI user and non-users), epidemiologists, computer scientists, the platforms that control the data, and the regulatory bodies that govern AI.

\bibliographystyle{plainnat}
\bibliography{references}

\end{document}